\documentclass[12pt,epsf,amstex]{article}
\usepackage [dvips]{graphicx}
\usepackage{amsmath}
\usepackage{amssymb}
\usepackage{epsfig}
\usepackage{verbatim}
\usepackage[usenames,dvipsnames,svgnames]{xcolor}

\addtocounter{secnumdepth}{1}
\font\capfont=cmbx12 at 50 pt 
\newbox\capbox \newcount\capl \def\a{A}
\def\docappar{\medbreak\noindent\setbox\capbox\hbox{%
\capfont\a\hskip0.15em}\hangindent=\wd\capbox%
\capl=\ht\capbox\divide\capl by\baselineskip\advance\capl by1%
\hangafter=-\capl%
\hbox{\vbox to8pt{\hbox to0pt{\hss\box\capbox}\vss}}}
\def\cappar{\afterassignment\docappar\noexpand\let\a }

\begin{document}

\newcommand{\ee}{{\rm e}}
\newcommand{\calT}{{\cal T}}
\newcommand{\calE}{{\cal E}}
\newcommand{\bey}{\boldsymbol{e}_y}
\newcommand{\bv}{\mathbf{v}}

\newcommand{\tq}{t_{\rm q}}
\newcommand{\tqm}{t_{\rm q}^*}
\newcommand{\tqs}{t_{\rm qs}}
\newcommand{\tmax}{t_{\rm q}^{\rm max}}
\newcommand{\tm}{t_{\rm m}}
\newcommand{\tauqs}{\tau_{\rm qs}}
\newcommand{\rhotwo}{\rho_2}
\newcommand{\rhoin}{\rho_{\rm in}}

\newcommand{\infy}{\inf_{\rule{0mm}{2.55mm}y}}

\newcommand{\bE}{\bar{E}}
\newcommand{\calH}{{\cal H}}
\newcommand{\calJ}{{\cal J}}
\newcommand{\calL}{{\cal L}}
\newcommand{\calM}{{\cal M}}
\newcommand{\calN}{{\cal N}}
\newcommand{\calW}{{\cal W}}
\newcommand{\hP}{\hat{P}}
\newcommand{\hPi}{\hat{\Pi}}
\newcommand{\sumn}{\sum_{n=1}^N}

\newcommand{\vecalpha}{\vec{\alpha}}
\newcommand{\vecg}{\vec{g}}
\newcommand{\vecp}{\vec{p}}

\newcommand{\tb}{\tilde{b}}
\newcommand{\tA}{\tilde{A}}
\newcommand{\tB}{\tilde{B}}
\newcommand{\tP}{\tilde{P}}
\newcommand{\tbeta}{\tilde{\beta}}
\newcommand{\tgamma}{\tilde{\gamma}}
\newcommand{\tcalM}{\widetilde{\cal M}}
\newcommand{\betast}{{\beta_*}}

\newcommand{\intp}{\int_{-\pi}^{\pi}\frac{\dd p}{2\pi}}
\newcommand{\intpone}{\int_{-\pi}^{\pi}\frac{\dd p_1}{2\pi}}
\newcommand{\intptwo}{\int_{-\pi}^{\pi}\frac{\dd p_2}{2\pi}}
\newcommand{\ointz}{\oint\frac{\dd z}{2\pi{\rm i}}}
\newcommand{\qext}{q_{\rm ext}}

\newcommand{\bO}{{\bf{O}}}
\newcommand{\bR}{{\bf{R}}}
\newcommand{\bS}{{\bf{S}}}
\newcommand{\bT}{{\bf{T}}}
\newcommand{\bn}{{\bf{n}}}
\newcommand{\br}{{\bf{r}}}
\newcommand{\bt}{\mbox{\bf t}}
\newcommand{\half}{\frac{1}{2}}
\newcommand{\thalf}{\tfrac{1}{2}}
\newcommand{\bsA}{\mathbf{A}}
\newcommand{\bsV}{\mathbf{V}}
\newcommand{\bsE}{\mathbf{E}}
\newcommand{\bsT}{\mathbf{T}}
\newcommand{\bse}{\mbox{\bf{1}}}

\newcommand{\invup}{\rule{0ex}{2ex}}
\newcommand{\rmc}{{\rm c}}

\newcommand{\dd}{\mbox{d}}
\newcommand{\p}{\partial}

\newcommand{\la}{\langle}
\newcommand{\ra}{\rangle}

\newcommand{\beq}{\begin{equation}}
\newcommand{\eeq}{\end{equation}}
\newcommand{\bea}{\begin{eqnarray}}
\newcommand{\eea}{\end{eqnarray}}
\def\lsim{\:\raisebox{-0.5ex}{$\stackrel{\textstyle<}{\sim}$}\:}
\def\gsim{\:\raisebox{-0.5ex}{$\stackrel{\textstyle>}{\sim}$}\:}

\numberwithin{equation}{section}   

\thispagestyle{empty}
\title{{\Large {\bf Density decay and growth of correlations\\[2mm] 
in the Game of Life}\\
\phantom{xxx}}}

\author{{\bf F.~Cornu and H.J.~Hilhorst}\\[5mm]
{\small Laboratoire de Physique Th\'eorique, UMR 8627}\\[-1mm]
{\small Universit\'e Paris-Sud and CNRS,}\\[-1mm]
{\small Universit\'e Paris-Saclay}\\[-1mm]
{\small B\^atiment 210, 91405 Orsay Cedex, France}\\}

\maketitle

\begin{abstract}
We study the Game of Life as a statistical system on an $L\times L$ 
square lattice with periodic boundary conditions. 
Starting from a random initial configuration 
of density  $\rho_{\rm in}=0.3$ we investigate the
relaxation of the density as well as the growth with time
of spatial correlations.
The asymptotic density relaxation is exponential with a characteristic time
$\tau_L$ whose system size dependence follows a power law
$\tau_L\propto L^z$ with $z=1.66\pm 0.05$ before 
saturating at large system sizes to a constant $\tau_\infty$.
The correlation growth is characterized by
a time dependent correlation length $\xi_t$
that follows a power law
$\xi_t\propto t^{1/z^\prime}$ with $z^\prime$ close to $z$
before saturating at large times to a constant $\xi_\infty$. 
We discuss the difficulty of determining the correlation length $\xi_\infty$
in the final ``quiescent'' state of the system. 
The decay time $\tq$ towards the 
quiescent state is a random variable;
we present simulational evidence as well as a heuristic
argument indicating that for large $L$ its distribution
peaks at a value $\tqm(L) \simeq 2\tau_\infty\log L$.

\end{abstract}


\noindent{\bf Keywords:} 
cellular automata; Game of Life; critical phenomena.

\newpage

\section{Introduction}
\label{sec:intro}

\cappar The Game of Life (GL) is a cellular automaton proposed in 1970 by 
Conway \cite{Berlekampetal82} and made popular by Gardner \cite{Gardner70}.
It evolves deterministically in discrete time $t=0,1,2,...$
according to the following rules.
On a two-dimensional square lattice,
at any instant of time $t$, each site $\br$
may be occupied or empty (occupation number $n_t(\br)=1$ or $n_t(\br)=0$).
The sites are often referred to as ``cells'' and the occupied and 
empty states are said to correspond to the cell being ``alive'' and ``dead,''
respectively.
The state of a site $\br$ at time $t+1$ follows deterministically
from its own state and the states of its eight neighbors 
(the Moore neighborhood) at time $t$;
the update rule is formulated with the aid of the
auxiliary quantity 
$S_t(\br)\in\{0,1,...,8\}$, defined as
the sum at time $t$ of the occupation numbers of the neighbors of $\br$. 
In terms of this sum Conway's update rule reads

--  if $S_t(\br)\neq 2,3$, then $n_{t+1}(\br)=0$~;

--  if $S_t(\br) = 2$, then $n_{t+1}(\br) = n_t(\br)$~; 

--  if $S_t(\br) = 3$, then $n_{t+1}(\br)=1$,\\
and is carried out synchronously for all sites.

The qualification ``totalistic'' is used to
indicate that the occupation numbers of the neighboring sites 
enter the update rule only through their sum.
The number of possible totalistic cellular automata based on the Moore
neighborhood is equal to $2^{18}$.
Several authors \cite{Wolfram84,WolframPackard85,Eppstein10}
have contributed to classifying the automata in this category.
Conway noticed that there is 
(i) a large ``supercritical'' subclass 
for which, typically, an initially localized set of living cells 
will progressively fill
the whole available lattice with living cells at some average density; and 
(ii) the complementary subclass for which no such explosive growth happens.

The GL update rule stated above stemmed from 
Conway's attempt to find the most interesting update rule.
In physical parlance, this meant that he
was trying to be as close as possible
to the critical line separating the two subclasses, while refusing to be
supercritical. 

Interest in the statistical properties of the GL goes back at least to the
work of Dresden and Wong \cite{DresdenWong75}, 
who adopted an analytical approach,
and that of Schulman and Seiden \cite{SchulmanSeiden78}, who
incorporated stochasticity in the rules of the game
and were followed therein by many later researchers.
Much interest in the GL was subsequently generated by
Bak {\it et al.} \cite{Baketal89}.
On the basis of their simulation of the effect
of small external perturbations repetitively applied to the GL
these authors claimed that due to ``self-organization'' 
\cite{Baketal87,Creutz92}
the GL is exactly {\it at\,} a critical point.
This idea has been advanced many times 
\cite{Bak92,AlstromLeao94,Creutz97,Fehsenfeldetal98,Turcotte99,%
Rozenfeldetal07},
either as a fact or as a hypothesis,
but was abandoned following the investigations of, in particular,
Bennett and Bourzutschky \cite{BennettBourzutschky91},
Hemmingsson \cite{Hemmingsson95}, 
Nordfalk and Alstr{\o}m \cite{NordfalkAlstrom96},
and Blok and Bergersen \cite{BlokBergersen97}.
Our work confirms, if that was still needed, that the
GL is subcritical; however, and as many authors have noted,
it is close to criticality. In this study
we investigate its near-critical properties
while staying strictly within the limits of the original GL:
we are interested in deterministic dynamics and
do not study any stochastic extensions of the GL, nor
subject it to external perturbations.

We will in this work study the GL on a periodic $L\times L$ lattice.
letting it start from an arbitrary random initial configuration.
It is well-known that under such circumstances the GL,
after a transient which may take thousands of time steps, 
enters a limit cycle, also referred to
as a ``quiescent'' or a ``stationary'' state. 
The quiescent state is composed 
of small independent ({\it i.e.}~nonoverlapping) groups of living cells
that we will call ``objects'' and  
that may be static or periodically oscillating 
(see {\it e.g.} \cite{Baketal89,Creutz97,Bagnolietal91}).
The vast majority of these objects belong to a dozen or so different types
with linear sizes in the range from two to five lattice units. 
The oscillators among them  
have almost all a period of two time units; 
oscillators of higher periodicities do exist but are statistically
insignificant. 
Related to the oscillators is the class of objects 
that are time-periodic modulo a translation in space.
In GL jargon these are referred to as ``spaceships,''
their simplest and foremost example being the ``glider.''
On dedicated websites 
(see {\it e.g.} \cite{GLsites})
a great deal of attention is paid to these special
objects. On a lattice with periodic boundary conditions
they may certainly occur during the relaxation process,
but their probability of survival into the quiescent state
is far too small for them to have an impact on the properties 
studied in this work. 
\vspace{2mm}

This paper is organized as follows.

In section \ref{sec:density} we consider the relaxation
of the density of living cells to its quiescent state value.
The relaxation curve is well-known\, \cite{Bagnolietal91}, 
but its asymptotic long time decay is subject to finite size effects that
have never been reported. The observation of these effects
requires strongly improved statistics, presented in section \ref{sec:density}. 
We extract from our simulations
the $L$ dependent decay time $\tau_L$ associated with the asymptotic
density decay. 
For $L\lesssim 60$ this decay time
appears to have the power law behavior $\tau_L\propto L^z$ with
$z\approx 1.66$, whereas after a crossover regime
it saturates for $L\gtrsim 180$ to a constant $\tau_\infty\approx 1800$.

In section \ref{sec:correlation} we consider
how density pair correlations develop in the course of time.
We are not aware of any earlier study of these time dependent correlations.
We establish that there is a correlation length $\xi_t$
that grows with time as $\xi_t \propto t^{1/z^\prime}$ 
and saturates for $t\gtrsim 8000$ to a constant $\xi_\infty\approx 50$.
Our value for $z^\prime$ is close to that of $z$ and we strongly suspect
that they are in fact one and the same exponent,
the difference being due to some unknown systematic bias in our analysis.
At the end of section \ref{sec:correlation}
we emphasize the difficulty of collecting statistics on the correlations in
the quiescent state.

In section \ref{sec:decaytimes}
we consider the probability distribution $P_L(\tq)$
of the decay times $\tq$ to quiescent state.
It appears that for large $\tq$, with very good accuracy, this distribution
decays exponentially with decay constant $1/\tau_L$.
A heuristic argument leads us to conclude that
for large $L$ the time $\tqm$ at which $P_L$ reaches its peak scales
as $\tqm(L) \propto \log L$. 
The predicted curve for $\tqm(L)$ is in excellent agreement with
the simulation data. 

In section \ref{sec:discussion} we 
discuss our results and 
compare them to related work in the literature.
We address several closely related points of interest and
also briefly return to the question of self-organized criticality.

\section{Density decay}
\label{sec:density}

\begin{table}[t]
\begin{center}
\begin{tabular}{||r|c|r||}
\hline
$L$ & $\rho_L(\infty)$ & $N$\phantom{000} \\
\hline
 128     &   0.028873 $\pm$  0.000009 & 100\,000\\
 144     &   0.028828 $\pm$  0.000027 &  10\,000\\
 160     &   0.028845 $\pm$  0.000022 &  10\,000\\
 176     &   0.028758 $\pm$  0.000025 &  10\,000\\
 192     &   0.028771 $\pm$  0.000012 &  10\,000\\
 208     &   0.028734 $\pm$  0.000017 &  10\,000\\
 224     &   0.028754 $\pm$  0.000009 &  10\,000\\
 240     &   0.028711 $\pm$  0.000018 &  12\,500\\
 256     &   0.028739 $\pm$  0.000008 &  40\,000\\
 512     &   0.028721 $\pm$  0.000006 &  40\,000\\
$\infty$ &   0.02872\phantom{0}  $\pm$  0.00001\phantom{0}  & \\
\hline
\end{tabular}
\caption{List of quiescent state densities $\rho_L(\infty)$
with $N$ the number of quiescent states having contributed to the average.
The last line is our extrapolation to infinite system size,
$\lim_{L\to\infty}\rho_L(\infty) \equiv \rho^*=0.02872(1)$.}
\label{tab:1}
\end{center}
\end{table}

\begin{figure}[t]
\begin{center}
\scalebox{.55}
{\includegraphics{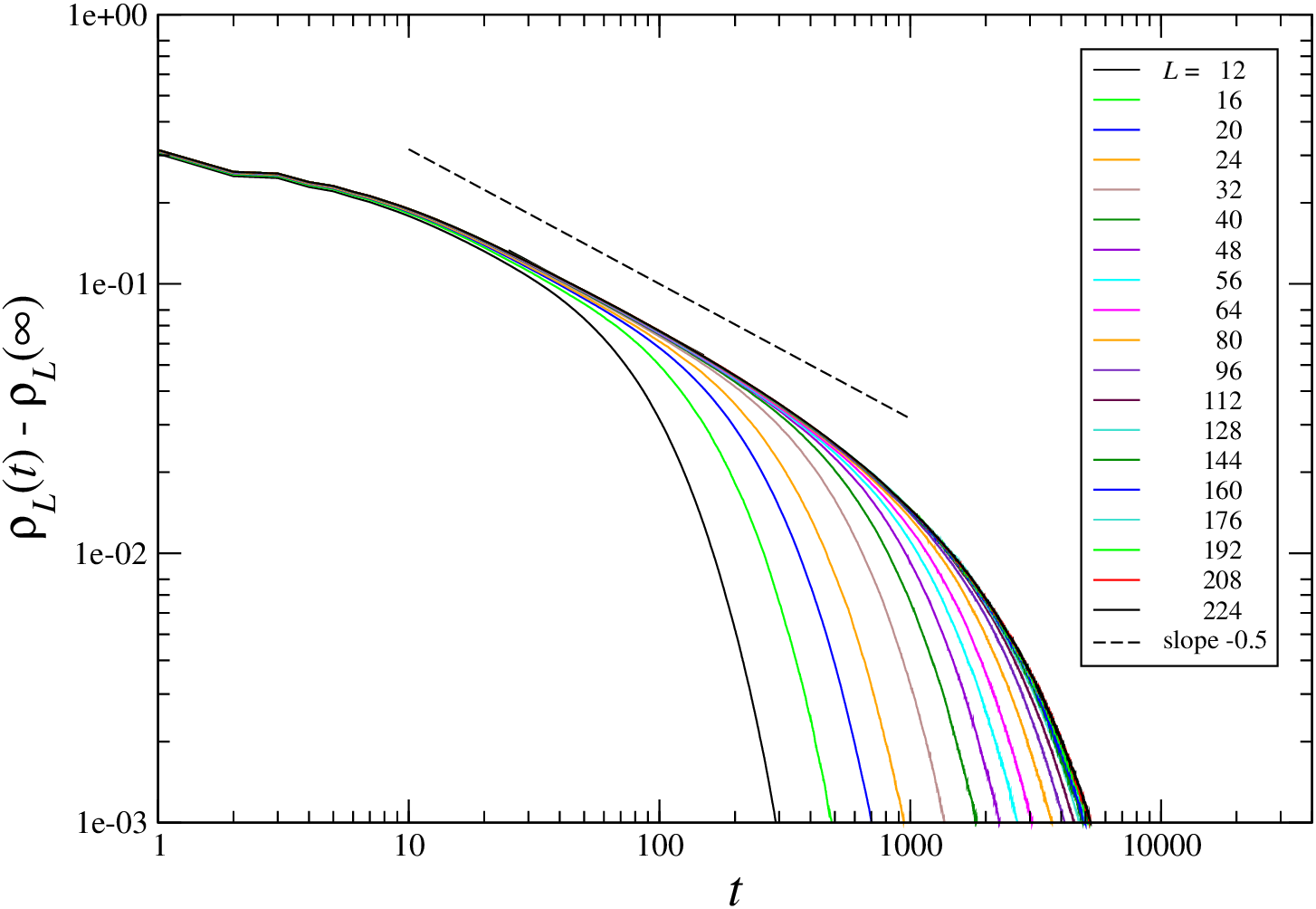}}
\end{center}
\caption{\small Density difference $\rho_L(t)-\rho_L(\infty)$
on a log-log scale for values of the system size $L$ that increase from left
to right.
For $L\gtrsim 100$ the curves become difficult to distinguish from their 
$L=\infty$ limit. The dashed line has slope $-0.5$.}
\label{fig:densrelax1}
\end{figure}

\begin{figure}[t]
\begin{center}
\scalebox{.55}
{\includegraphics{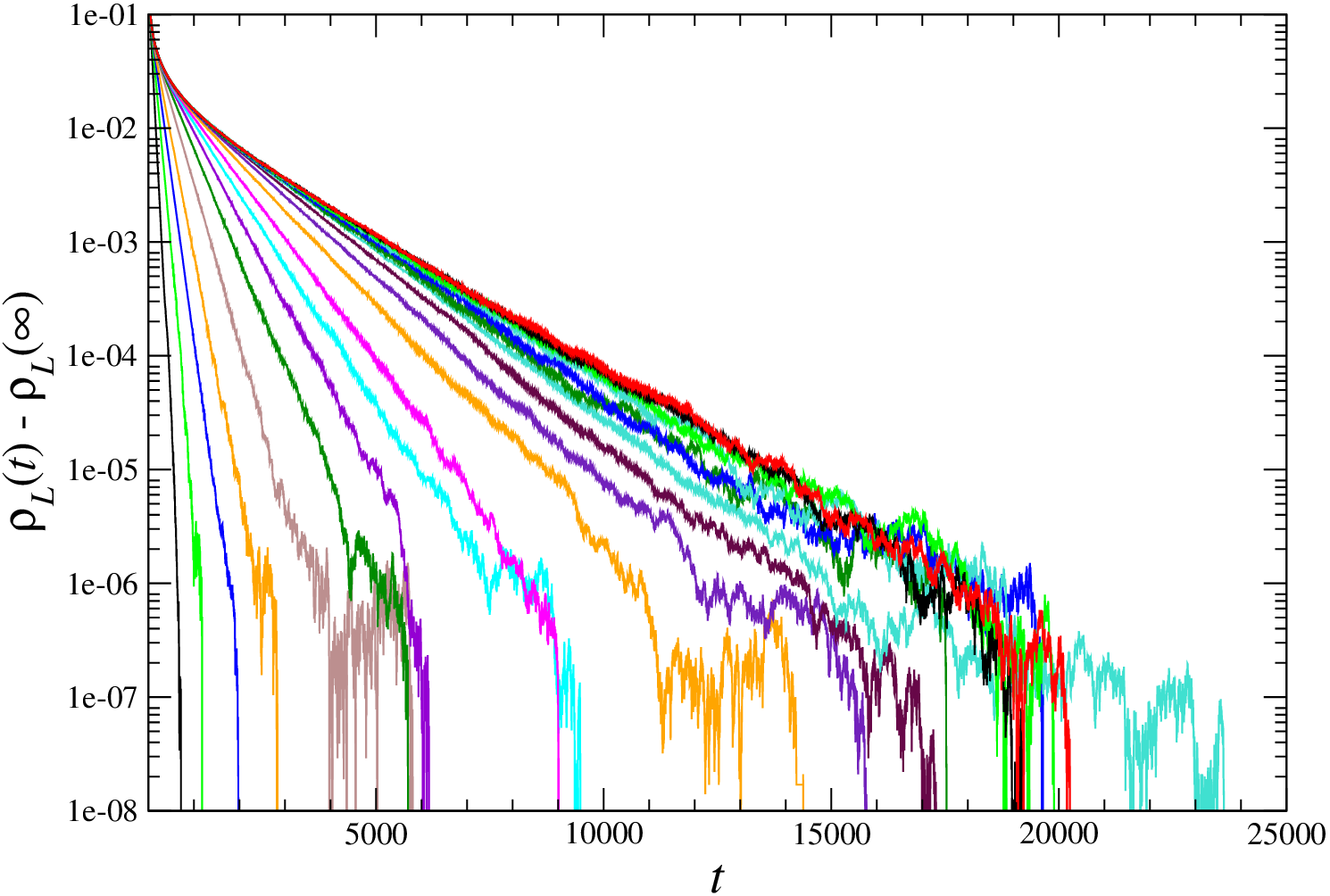}}
\end{center}
\caption{\small Density difference $\rho_L(t)-\rho_L(\infty)$
on a linear-log scale. 
Note that the vertical scale here goes down to much smaller values than
in figure \ref{fig:densrelax1}. The curves are for the same values of $L$
with the same color code as in figure \ref{fig:densrelax1}.
Curves for larger values of $L$ overlap among themselves and
with the limit curve, and are not shown.}
\label{fig:densrelax2}
\end{figure}

\begin{figure}[t]
\begin{center}
\scalebox{.55}
{\includegraphics{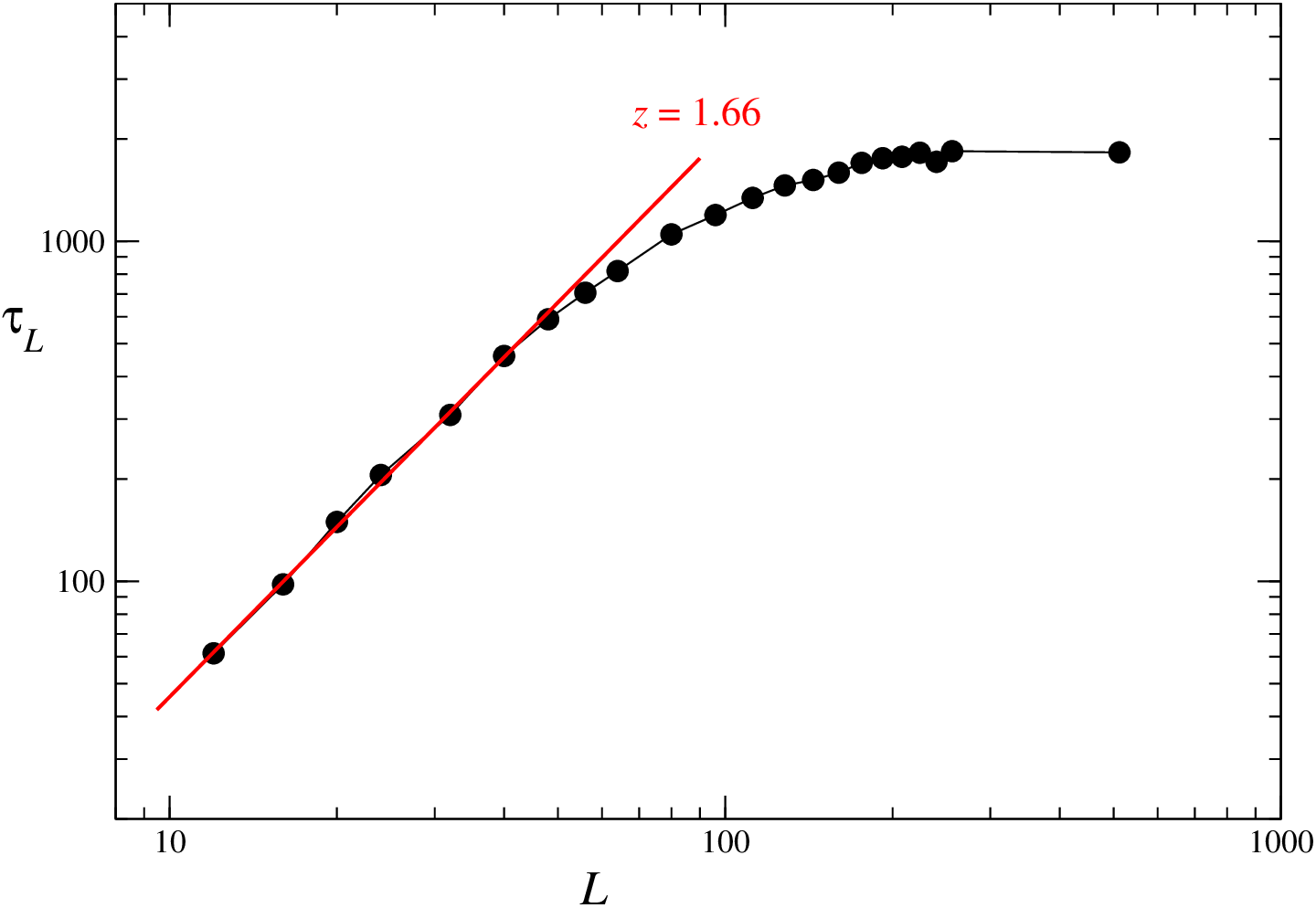}}    
\end{center}
\caption{\small Data points: the relaxation times $\tau_L$  
extracted from figure \ref{fig:densrelax2} shown on a log-log scale. 
Red line: best linear fit, having a slope $1.66$.}
\label{fig:tauL}
\end{figure}

\begin{figure}[t]
\begin{center}
\scalebox{.55}
{\includegraphics{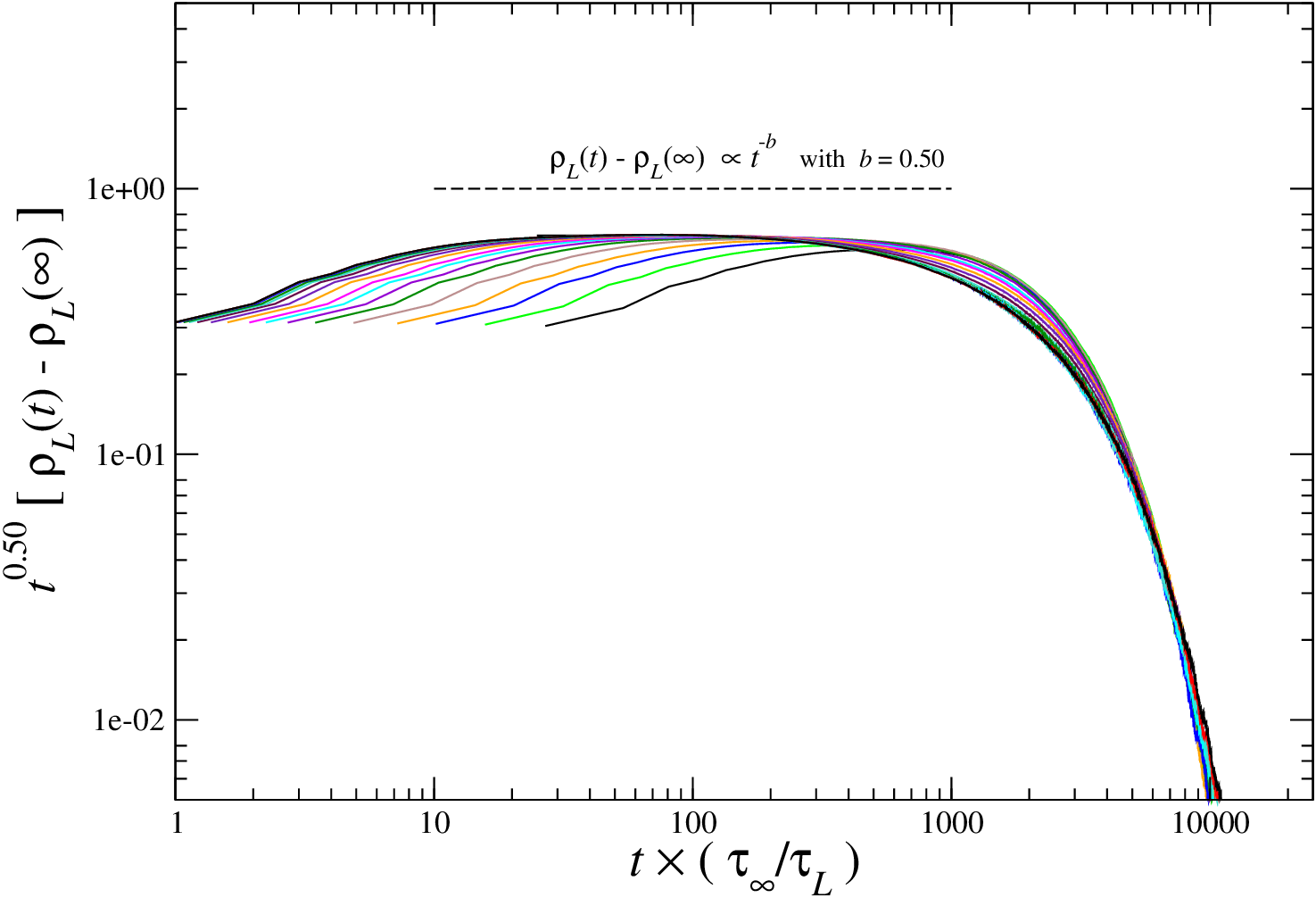}} 
\end{center}
\caption{\small The curves of figure \ref{fig:densrelax1} 
multiplied by $t^{0.50}$ and with the time scaled such that their
exponential tails coincide. The values of $L$ and the color code are
as in figure \ref{fig:densrelax1}. The power law regime appears as a plateau
(dashed line).}
\label{fig:densrelax3}
\end{figure}

We have simulated the time evolution of the GL
on an $L\times L$ lattice
for linear system sizes up to $L=512$.
The system was started
in a random initial configuration (``soup'') 
of density $\rho_{\rm in}=0.3$,
meaning that each site was independently alive with
probability $\rhoin$ or dead with probability $1-\rhoin$.
We then observed the decay with time of the density 
$\rho_L(t) \equiv L^{-2}\sum_{\br}\langle n_t(\br)\rangle$, 
where the angular brackets $\langle\ldots\rangle$
stand for the average over the random initial configurations.
Since we wish to analyze the decay of the density 
{\it difference\,} $\rho_L(t)-\rho_L(\infty)$,
which is analogous to an order parameter,
we have in our simulations first determined the
$L$ dependent quiescent state density $\rho_L(\infty)$.

\subsection{Quiescent state density $\rho_L(\infty)$}
\label{sec:asptdensity}

Table \ref{tab:1} shows our simulation results for 
the quiescent state density $\rho_L(\infty)$
for system sizes from $L=128$ up, together with the number $N$ of quiescent
states that contributed to each average.
The error bars were obtained from the dispersion among ten subsets
of quiescent states.
The convergence to the infinite lattice limit
$\rho^*\equiv\lim_{L\to\infty}\rho_L(\infty)$
seems to be at least exponential in $L$, but it is 
hard to ascertain its rate.
The last line of table \ref{tab:1}  lists what we feel is the best possible
estimate, 
\beq
\rho^* = 0.02872 \pm 0.00001,
\label{xrhostar}
\eeq
which is compatible with, and slightly more accurate than, 
the best earlier determinations \cite{Bagnolietal91,GibbsStauffer97}.

\subsection{Density decay at intermediate times}
\label{sec:intermdecay}

In figure \ref{fig:densrelax1}, using now the values of the 
$\rho_L(\infty)$ determined above, we show in a log-log plot
the decay curves of the density difference $\rho_L(t)-\rho_L(\infty)$
for various values of $L$.
The most prominent feature is that
in the large lattice limit the decay curves
clearly converge to a limit curve
$\rho(t)\equiv\lim_{L\to\infty}\rho_L(t)$. 
There appears to be an interval of almost two decades of intermediate times
where the limit curve is close to linear, signaling 
a power law relation 
\beq
\rho_L(t)-\rho_L(\infty)\propto t^{-b}, \qquad 10 \lesssim t \lesssim 1000.
\eeq
The slope of the dashed straight line, drawn for comparison,
shows that $b \approx 0.5$.

The steepening of the limit curve at times $t\gtrsim 2000$ 
indicates the crossover from power law to exponential decay. 
The GL is therefore subcritical: had it been critical, the power law would
have held up to increasingly longer times for increasing $L$.



Bagnoli {\it et al.} \cite{Bagnolietal91}  
provide essentially the same limit curve,  
which they obtained for a lattice size of $320 \times 200$ sites,
but plot the density, instead of  
the density difference, as a function of time.
For intermediate times such an analysis 
leads to a good linear fit 
$\rho(t) \propto t^{-\tb}$, but with a different exponent $\tb$ 
about equal to $\tb\approx 0.3$; 
Garcia {\it et al.} \cite{Garciaetal93} in later work
report $\tb = 0.39 \pm 0.04$.
Neither work discusses the finite size behavior of 
this curve, which will be our next object of investigation. 


\subsection{Density decay for $t\to\infty$}
\label{sec:asptdecay}

Figure \ref{fig:densrelax1} shows that the smaller the lattice size $L$,
the earlier the exponential decay sets in.
We now proceed to discuss these finite size effects.

Figure \ref{fig:densrelax2} presents the 
same relaxation curves as figure \ref{fig:densrelax1}, 
but in a log-linear plot which turns 
the exponential tails of the decay curves into straight lines.
It appears that we have
\beq
\rho_L(t)-\rho_L(\infty) \propto \ee^{-t/\tau_L}
\eeq
without any extra multiplicative power of time on the RHS.
Note that in figure \ref{fig:densrelax2} 
the differences $\rho_L(t)-\rho_L(\infty)$ go down 
to values as low as $10^{-6}$, as compared to only $10^{-3}$ 
in the time regime shown in figure \ref{fig:densrelax1}.
In figure \ref{fig:densrelax2} the intermediate power law decay has 
become nearly indistiguishable in
the extreme upper left corner of the graph.
The limiting curve is truly exponential
only when $\rho(t)-\rho(\infty)$ is of the order of a few thousandths.
In order that we obtain good statistics for such small density differences
the curves for the largest values of $L$
required the largest computational effort (see table \ref{tab:1}): 
those for $L\geq 144$
are averages over the time evolution of a number $N$ of initial configurations
between $10\,000$ and $40\,000$.
The curves for the smaller $L$ are averages
over at least $100\,000$ initial configurations.

It is easy to extract the asymptotic decay
time $\tau_L$ for each lattice size $L$
from the linear part of the corresponding decay curve.
The $\tau_L$ are shown in a log-log plot in figure \ref{fig:tauL}.
For moderate $L$ they go up as a power law of $L$,
but then saturate at a value $\tau_\infty\approx 1800$.
Numerically we find 
\beq
\tau_L \simeq \left\{
\begin{array}{ll}
A L^z, & A=1.03, \quad z=1.66 \pm 0.04,\\[2mm]
\tau_\infty = 1800 \pm 50,\phantom{XXX}& L\to\infty,\\
\end{array}
\right.
\label{xtauL}
\eeq
where $z$ is the dynamical exponent.
The individual error bars (not indicated) of each data point in figure
\ref{fig:tauL} is chiefly due to the choice of the interval for the linear fit
in figure \ref{fig:densrelax2}.
The small scatter in the set of data points is representative for these
individual errors.
The error bar $\pm 0.04$ in $z$ is due to the variations in slope of the red
line still compatible with the data points. The error in $\tau_\infty$ is
based on what seems a reasonable extrapolation. 

Figure \ref{fig:tauL} shows that deviations from the power law first
begin to occur, roughly, for $L \gtrsim 60$.
We expect that this length scale also
corresponds to a spatial correlation length and will investigate 
that idea in section \ref{sec:correlation}. 

For later reference we show in figure \ref{fig:densrelax3} 
the curves of figure \ref{fig:densrelax1}
multiplied by the power $t^b$ (taking $b=0.50$)
and with their time scaled by $\tau_L$.
As a consequence,
the power law regime appears as a plateau
and the exponential tails coincide. 
In the regime of scaled times between roughly $500$ and $5000$
corrections to this finite size scaling are clearly visible;
we have not investigated these any further. 

\section{Time dependent correlation length}
\label{sec:correlation}

\begin{figure}[t]
\begin{center}
\scalebox{.55}
{\includegraphics{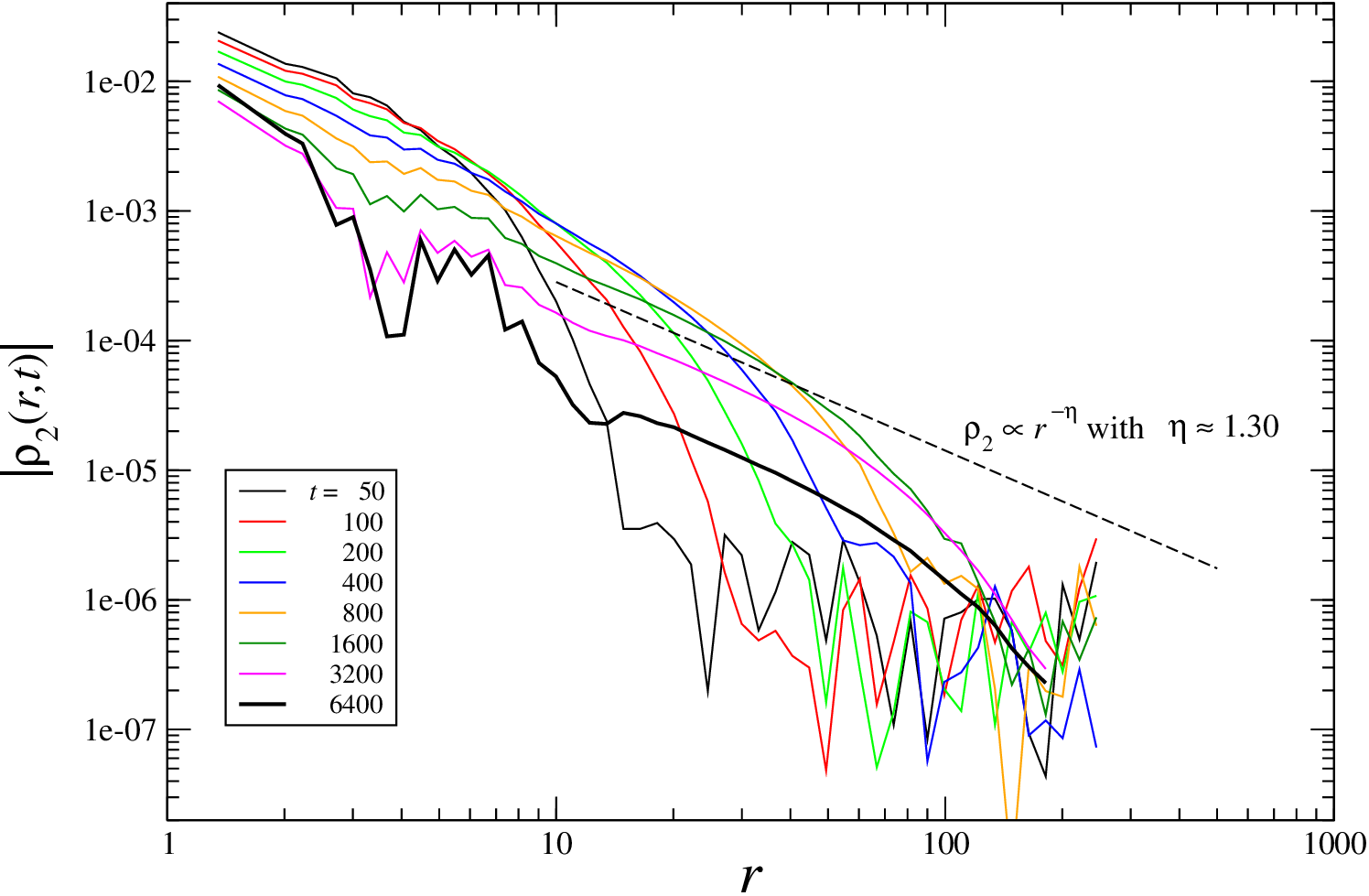}} 
\end{center}
\caption{\small Truncated pair density $\rhotwo(r,t)$
as a function of the radial distance $r$.
The curves are for times $t=50,100,\ldots,6400$.
All simulations are for a lattice of $512  \times 512$ sites.
The curves for the largest times, $t=3200$ and $t=6400$, are averages over
$100\,000$ and $50\,000$ configurations, respectively.
Those for the smaller times are each on $5000$ configurations.
The dashed line has slope $-1.30$ and indicates the power law behavior
discernible in an intermediate time regime of each curve before it 
steepens.}
\label{fig:corrgrowth1}
\end{figure}

We wish to interpret the length scale identified above as a correlation length 
and to that end we now investigate the correlations between
the occupation numbers on sites a distance $r$ apart.
To the best of our knowledge, such correlations have not been studied before. 

In the initial configuration the site
occupation numbers of distinct sites are uncorrelated.
For $t>0$ correlations will develop and we
expect that there will be a time dependent correlation length 
$\xi_t$. The system being subcritical, this
length must in the large time limit
necessarily saturate at some finite constant value $\xi_\infty$.

For the simulations that follow we would ideally like to work on an infinite
lattice. In practice we have worked on a $512 \times 512$ lattice,
whose linear size is considerably larger than the range of the expected
correlations.
We will conclude {\it a posteriori\,} that for our purposes this lattice size
is practically the ``thermodynamic limit.''

We have determined the time dependent 
truncated pair density function 
\beq
\rhotwo(r,t) \equiv \langle n_t(0)n_t(\br) \rangle - \langle n_t(0) \rangle^2
\label{dn2rt}
\eeq
where, as before, $\langle\ldots\rangle$ denotes the average over 
the random initial ensemble at density $\rho_{\rm in}=0.3$.
The notation $\rhotwo(r,t)$ in (\ref{dn2rt}) is meant to include 
both a translational average and
an average over the spatial annulus with $|\br| = r$. 

 
Our interest is in the large $r$ behavior of the pair density
and we have checked that circular symmetry sets in quickly for
distances $\gsim 10$ lattice units.
To represent $\rhotwo(r,t)$ we divided the axis of the
variable $\log r$ into equal-size bins centered at $r=r_k$
where $\log r_k = k/10$ for $k=0,1,2,\ldots$.
This, combined with the annular average, leads to
better statistics%
\footnote{Some of the bins with small $k$ values are empty, but
this is of no importance for our analysis of the large $r$ behavior.} 
for large values of $r$.
We have determined $\rhotwo(r,t)$
for all $k$ such that $r_k \leq 180$.
Figure \ref{fig:corrgrowth1} shows our
raw data for the pair correlation $\rho_2(r,t)$,
plotted as a function of $r$ 
for the geometric sequence of times
$t=50,100,200,...,6400$.  
The figure first of all corroborates the expected
phenomenon that the range of the correlations
increases with time.
It warrants numerous comments, that we will
make in the four following subsections.

\subsection{Correlation decay at intermediate and short distances}
\label{sec:intermcorrdecay}

In an intermediate spatial range that depends on $t$
the curves show the power law behavior  
$\rho_2(r,t) \propto r^{-\eta}$ with $\eta = 1.30 \pm 0.10$.
Although this intermediate range does not cover more than a decade, 
its existence again shows that the GL is not far from criticality.


The values of $\rho_2(r,t)$
for $r\lesssim 10$ represent short range correlations and
are outside of our focus of attention,
but we must nevertheless comment on them.
The precise shape of $\rhotwo(r,t)$ in this spatial region is
definitely dependent on our choice of binning.
The truncated pair density may, and in fact does have zeros on the $r$ axis.
For $t=50$ a negative dip in the correlation occurs around $r=27$
and is followed by at least two further oscillations.
As $t$ increases, 
this first dip moves to higher $r$, such that for $t=400$ it is
located at $r=95$ and for $t=800$ it has moved out of the range of 
$r$ covered by our simulations.
However, the existence of dips at larger $r$ may still reflect upon the $r$
dependence of the tail visible in the simulation.

For our longest times, $t=3200$ and $t=6400$,
the short range structure in $\rhotwo(r,t)$ is seen to strongly increase.
We comment on this in subsection \ref{sec:latetimes}.

\subsection{Correlation decay for $r\to\infty$}
\label{sec:corrlength}

\begin{figure}[t]
\begin{center}
\scalebox{.55}
{\includegraphics{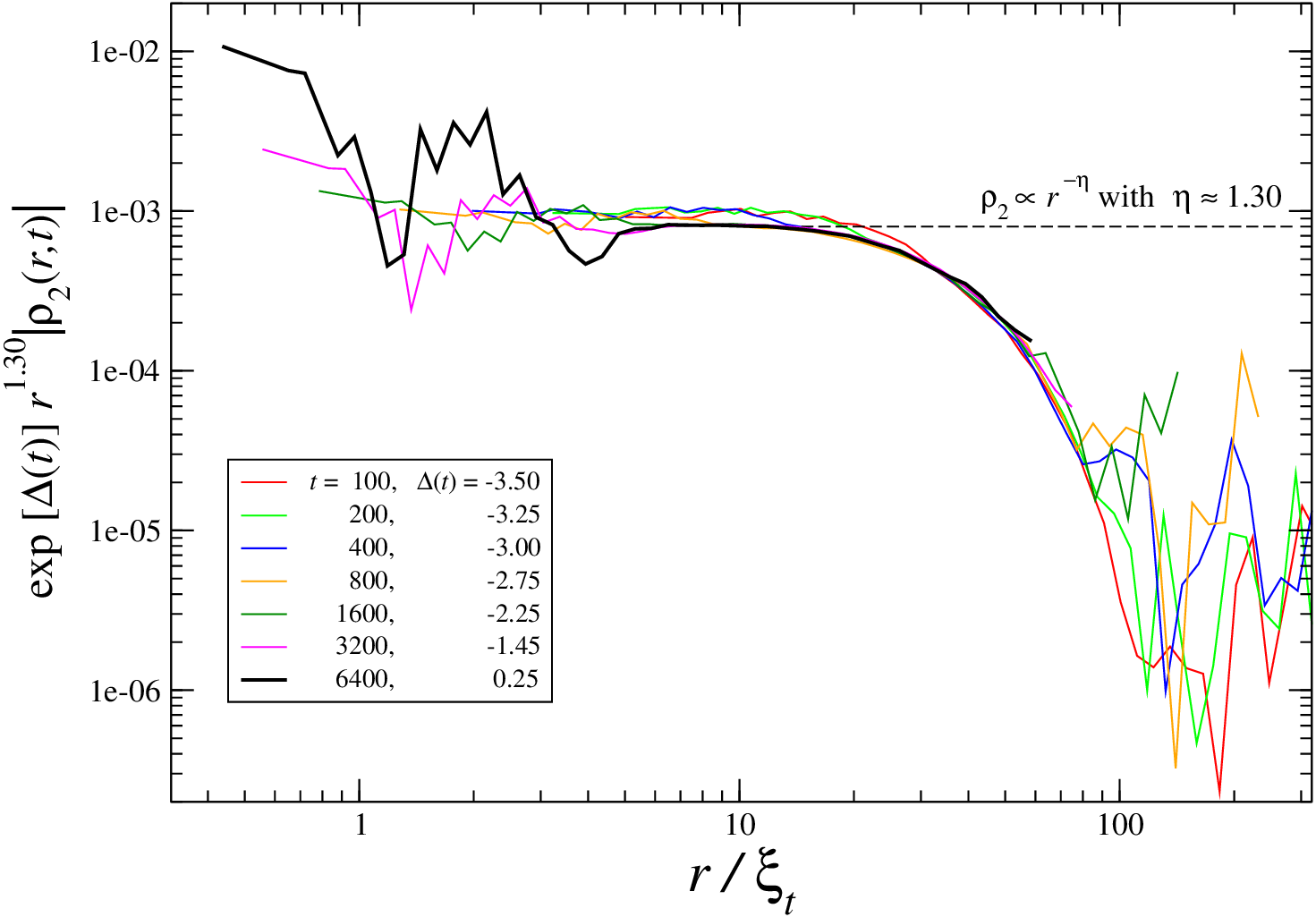}} 
\end{center}
\caption{\small The curves of figure \ref{fig:corrgrowth1} 
multiplied by $r^{\eta}$ and shifted by $\Delta(t)$,
as a function of the scaled coordinate $r/\xi_t$. 
This makes them coincide in the region
$20 \lesssim r/\xi_t \lesssim 80$
where their spatial decay begins to deviate from the power law $r^{-\eta}$.
}
\label{fig:corrgrowth2}
\end{figure}

Figure \ref{fig:corrgrowth2} shows a log-log plot of the data
(except the $t=50$ curve) of figure \ref{fig:corrgrowth1} 
after they have been subjected to 
a scaling similar to, although more complicated than, the one that led to
figure \ref{fig:densrelax3}.
The scaling here consists of the following three operations:

${}$\phantom{ii}(i) Multiplication by the power $r^{1.30}$, which results in 
the power law regimes appearing as plateaus;
the plateau is not very well developed at short times but becomes better
visible at later times. 

${}$\phantom{i}(ii) Multiplication by a constant $\ee^{\Delta(t)}$
chosen such that all plateaus are at the same
common level indicated by the dashed line in the figure.\\
\noindent The ``shift'' $\Delta(t)$ is determined only up to a constant;
we have arbitrarily set $\Delta(5600)=0$ (curve not shown).
This shift describes the decay with time of the amplitude
of the correlation.
One might have expected
that $\ee^{-\Delta(t)}$ would be proportional to $\rho^{-2}(t)$,
which in many cases in statistical physics is the natural normalization factor
for the pair density; this simple normalization does not, however,
appear to hold here.
We find that for $400 \lesssim t \lesssim 2400 $
the shift $\Delta(t)$ is roughly linear in $t$. 

(iii) Rescaling of the abscissa from $r$ to $r/\xi_t$,
with the $\xi_t$ chosen such that 
for all curves their initial deviations from the plateau value 
occur at the same rescaled time. This determines the ratios of the $\xi_t$.
We observe that there is a good scaling; its
mathematically expression is
\beq
\rhotwo(r,t) \simeq \frac{\ee^{-\Delta(t)}}{r^\eta}\,
{\cal G}\left(r/\xi_t\right),
\label{rhotwoscaling}
\eeq
valid in the time regimes of the power law and of its crossover
to a steeper decay.
It appears that in the region where the deviations from power law 
behavior first occur, {\it i.e.} for $20 \lesssim r/\xi_t \lesssim 80$,
the scaling function 
is best approximated not by an exponential but by 
${\cal G}(u) = \mbox{cst}\times u\,\ee^{-\kappa u}$.
We have fixed the scale of the abscissa in figure
\ref{fig:corrgrowth2} by the choice $\kappa=1$. 

\begin{figure}[t]
\begin{center}
\scalebox{.55}
{\includegraphics{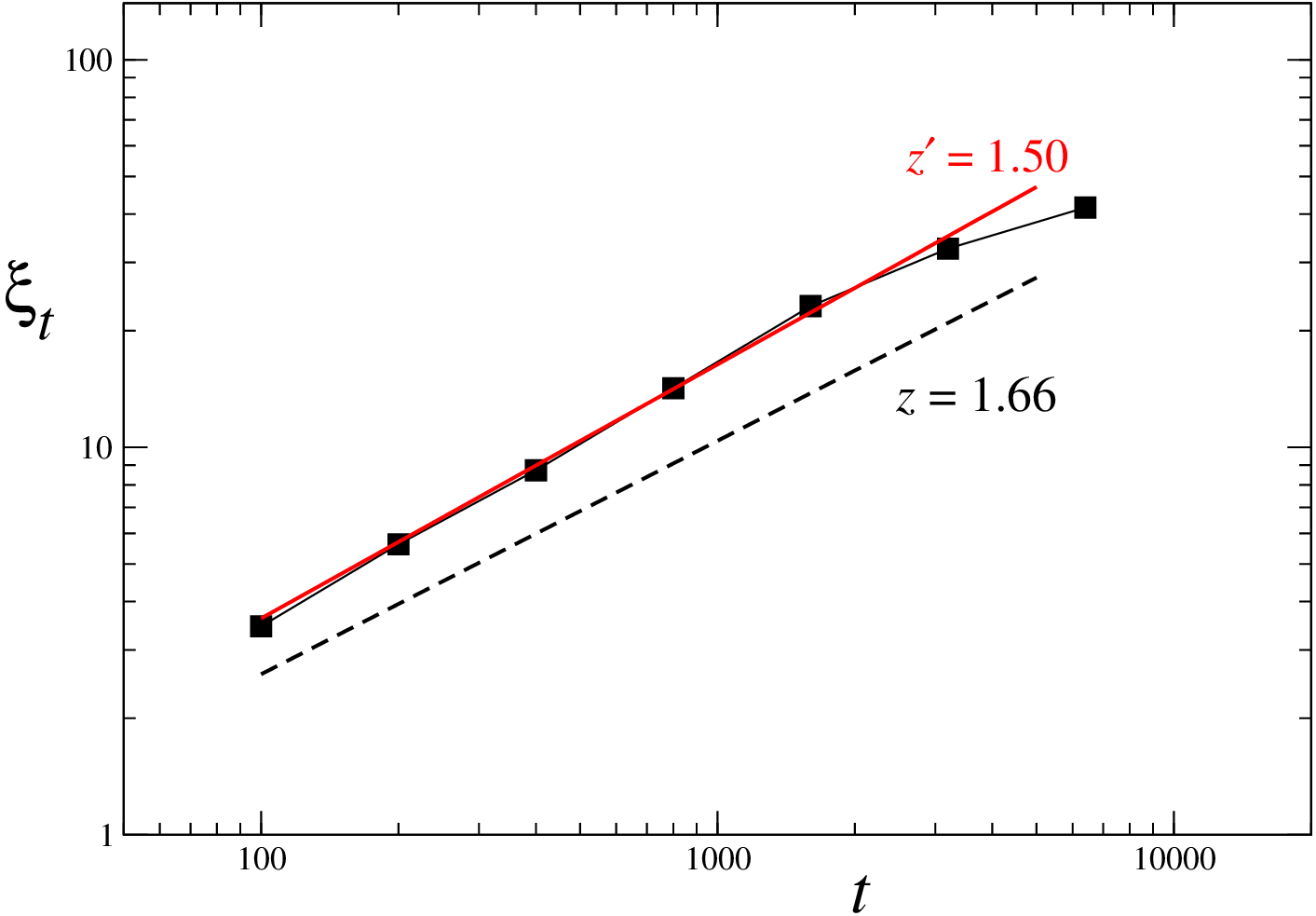}} 
\end{center}
\caption{\small Data points: the correlation length $\xi_t$
as a function of time $t$ on a log-log scale.
The red line is the best linear fit and has 
a slope $1/z^\prime$ with $z^\prime=1.50 $,
the dashed black line, shown for comparison, has slope $1/z$ with
$z=1.66$ as determined in section \ref{sec:asptdecay}.}
\label{fig:xit}
\end{figure}

The resulting values of $\xi_t$ 
are shown on a log-log scale
in figure \ref{fig:xit}. Over most of the range shown 
there is a linear dependency, indicating the power law relation
$\xi_t \propto t^{1/z^\prime}$.
Deviations from this relation 
begin to occur only for $t=3200$ and become more apparent for $t=6400$.
The saturation of $\xi_t$ at some finite value $\xi_\infty$, although only
slow, seems clearly indicated; numerically we estimate
\beq
\xi_t \simeq 
\left\{
\begin{array}{ll}
B\,t^{1/z^\prime}, &  B=0.12, \qquad z^\prime=1.5 \pm 0.1,\\[2mm]
\xi_{\infty}=50 \pm 10,   \phantom{XX} & t\to\infty.
\end{array}
\right.
\label{xxit}
\eeq
Figure \ref{fig:xit} shows for comparison also the straight line 
corresponding to the exponent value $z=1.66$ 
found in section \ref{sec:density}.
The two slopes are visually close to parallel, even though
our estimates for the statistical errors in $z$ and $z^\prime$ 
fall short of overlapping.

The essence of our remarks below equation (\ref{xtauL}) 
about the error bar estimation 
remains valid here.
In the absence of strong indications to the contrary,
we adhere to the simplest scenario in which at each time $t$ there is
only a single length scale, so that $z^\prime=z$.
Our error bars reflect the statistical errors but
do not take into account any systematic effects
that there might be.
We tentatively attribute the difference between the
two values to a small but unknown systematic bias that we believe
most likely affects the analysis of the
correlation function, which is less straightforward than
that of the density decay.

We note, finally, that
the limit value $\xi_\infty$ in figure \ref{fig:xit}
cannot be estimated with the same accuracy
as $\tau_\infty$ in figure \ref{fig:tauL}.

\subsection{Correlations upon the approach of the quiescent state}
\label{sec:latetimes}

\begin{figure}[t]
\begin{center}
\scalebox{.55}
{\includegraphics{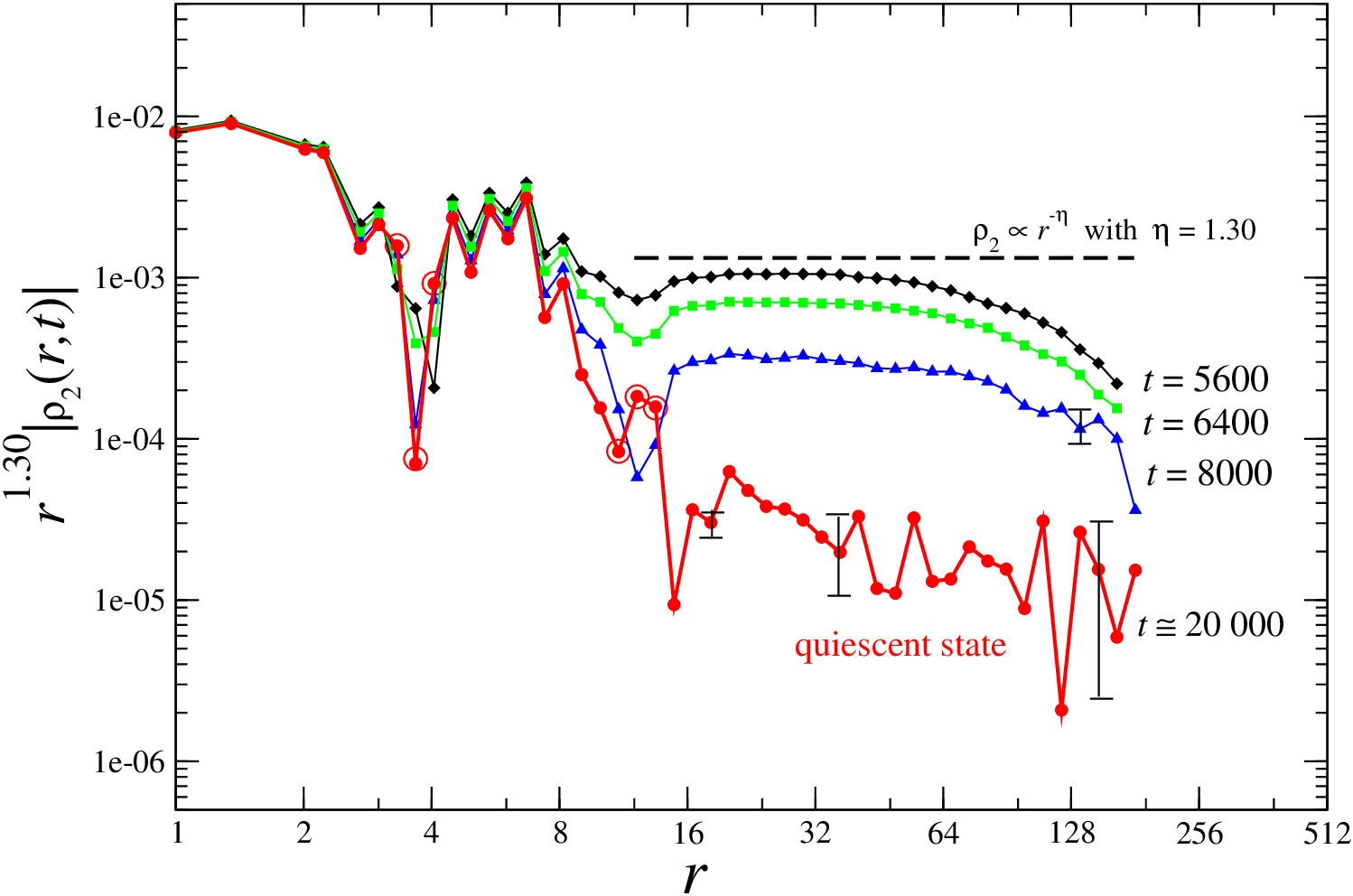}} 
\end{center}
\caption{\small Truncated pair density function for four late times.
The time $t=20\,000$ corresponds practically to the quiescent state. 
Error bars are discussed in the text;
a few typical ones have been indicated.
For the quiescent state (red curve) the six data points where
$\rho_2$ is negative have been encircled. 
All runs were performed for lattice size $L=512$.}
\label{fig:corrgrowth3}
\end{figure}

We now discuss the evolution of the pair correlation at 
late times, when the system approaches its quiescent state.

For the $512 \times 512$ 
(essentially infinite) 
lattice 
the time decay of the density difference $\rho_L(t)-\rho_L(\infty)$
starts its crossover to exponential around $t\approx 1000$,
but becomes truly exponential only for times as late as
several thousand time units. The density 
$\rho_L(t)$ of living cells is then
only a few thousandths above the quiescent state value $\rho^*=0.02872$.

The nature of the quiescent state has been recalled in the introduction.
The small static and periodic objects of which it consists 
cannot overlap -- if two of them were to overlap, they would interact and
transform into something else. Therefore the constituents of
the quiescent state act as ``hard objects''
with a typical diameter of the order of a few lattice units.
This is a new short distance length scale, which changes the 
structure of the pair correlation.

Figure \ref{fig:corrgrowth3}
shows the pair density (multiplied by the power law $r^{1.30}$)
at four different times:
 $t=5600,\, 6400,\, 8000,$ and $ 20\,000$.
The curves are averages over a number $N$
of independent systems equal to
$N=35\,000,$\, $50\,000,$\, $20\,000,$ and $35\,000$, 
respectively.
At time $t=20\,000$ (red curve)
virtually all systems have reached the quiescent state.

All these runs were performed for lattice size $L=512$.
Error bars were determined from the variance of ten subgroups of results.
As expected, all error bars increase with $r$.
For $t=5600$ and $t=6400$ they remain nevertheless at most of the order 
of the symbol size over the full range shown.
For $t=8000$ they begin to considerably exceed the
symbol size when $r\gtrsim 100$.

When the time $t$ is further increased, the curve $\rho_2(r,t)$
seemingly becomes a chaotic function of $r$.
This appearance is due to two very different effects
which we will discuss for $t=20\,000$.
\vspace{2mm}

${}$\phantom{i}(i)
First, for $r \lesssim 20$ the error bars in the $t=20\,000$ curve
are still small, and 
what looks like a random curve is actually a reproducible structure,
generated by the appearance of the new short range length scale.
This is well illustrated by the phenomenon that we see happening near $r=12$ 
for late times. Near this point, the black and green curves 
begin to develop a dip that gradually deepens (blue curve).
Beyond a certain time, $t \approx 9000$, 
the pair density at $r=12$ goes negative,
as signaled by the encircled data points of the red curve.%
The associated oscillating behavior in space is analogous to
that of the pair correlation in a dense liquid; in the present case
the atoms are not the individual living cells, but the elementary static
or periodic objects into which they have aggregated.
\vspace{2mm}

(ii) Secondly, for $r \gtrsim 20$, the randomness in the $t=20\,000$ 
curve is due to the difficulty of collecting good statistics. 
In this regime the error bars, some of which have been indicated, 
increase hugely with $r$ and this part of the curve is not reproducible.
The root cause is again the aggregation of living cells into 
a few types of larger objects; this reduces the effective number of degrees of
freedom without reducing the computational requirements,
and hence makes averaging less efficient.
\vspace{2mm}

The question of whether the GL quiescent state is critical, {\it i.e.,} has
infinite correlation length, is of definite interest.
It is from this state that Bak {\it et al.} \cite{Baketal89}
and later authors start in their
attempts to show or disprove that the
GL is self-organized critical.
The present work provides strong indication that for $t\to\infty$ the
GL correlation length $\xi_t$ tends to a {\it finite\,} value $\xi_\infty$. 
The state of affairs described above
makes clear, however, that it is 
very hard to track $\xi_t$ down in a simulation all the way to
the quiescent state, or, in practice, to  $t=20\,000$.
We briefly return to related questions in section \ref{sec:discussion}. 
 
\section{Decay time distribution}
\label{sec:decaytimes}

\begin{figure}[t]
\begin{center}
\scalebox{.55}
{\includegraphics{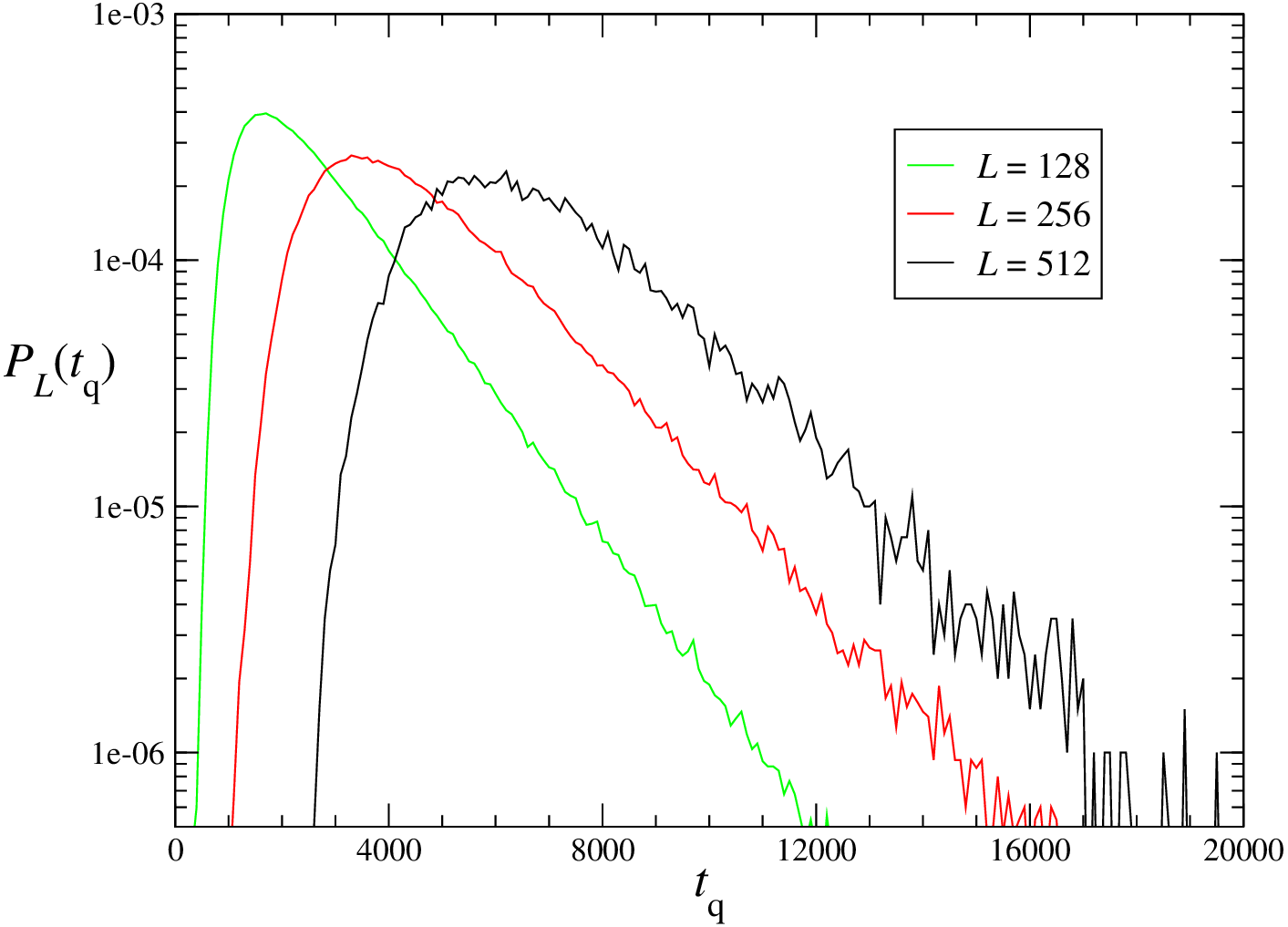}} 
\end{center}
\caption{\small Probability distribution 
$P_L(\tq)$ of the decay time $\tq$ to the quiescent state in 
  systems of linear size $L=128,\, 256$, and $512$. The exponential decay of the
long time tail is manifest.}
\label{fig:decay1}
\end{figure}

\begin{figure}[t]
\begin{center}
\scalebox{.55}
{\includegraphics{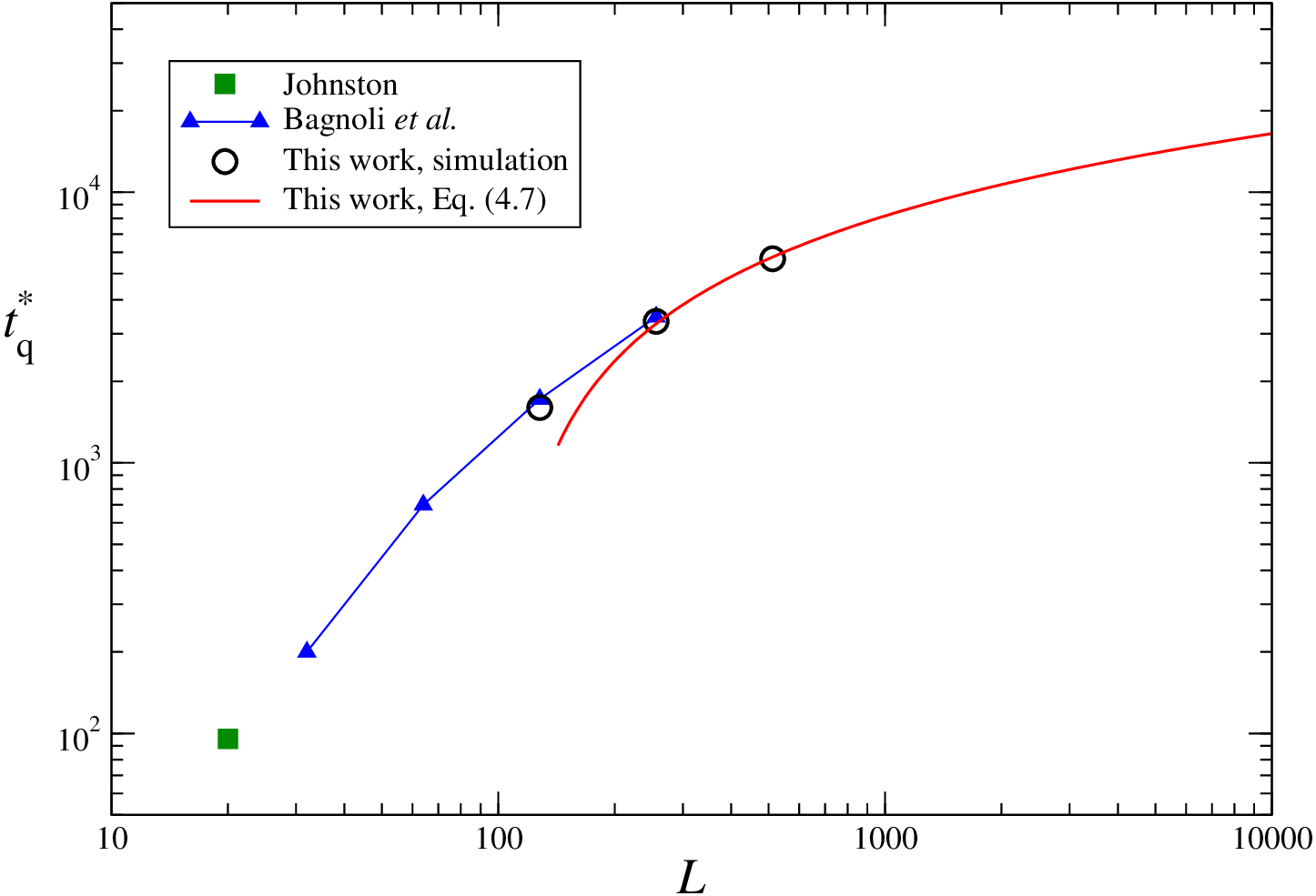}} 
\end{center}
\caption{\small Time $\tqm$ for which $P_L(t)$ has a maximum,
as a function of $L$.
The square data point for $L=20$ is due to Johnston
\cite{Johnston09}, the triangular ones
for $L=32,\, 64,\, 128$, and $256$ are due to Bagnoli {\it et al.}
\cite{Bagnolietal91} and have been connected to guide the eye, 
and the circular ones,
for $L=128,\, 256$, and $512$, were obtained in this work.
The red line is the large $L$ expansion [equation (\ref{soltqm})]
of our heuristic theory.}
\label{fig:decay4}
\end{figure}

Let $\tq$ denote the random instant of time (the ``decay time'')
at which the system reaches its
quiescent state. This random variable
is of course determined by the random initial configuration
and we will denote its distribution by $P_L(\tq)$.
In this section we discuss how the dynamic exponent $z$ appears in this
distribution.

\subsection{Simulation data}
\label{sec:decaytimessimul}

For system sizes $L=128,\, 256$, and $512$ 
we determined the distribution $P_L(\tq)$ from an ensemble of
$900\,000$, $150\,000$, and $20\,000$ initial states, respectively.%
\footnote{To plot $P_L(\tq)$ we divided the abscissa into time intervals
$[100(\ell-1),100\ell]$ with $\ell=1,2,3,\ldots$. During the simulation we
determined for each time interval the minimum and the maximum
number, $N_{\min}(\ell)$ and  $N_{\max}(\ell)$, respectively,
of living cells that occurred. When at the end of the $(\ell+1)$th interval
we found that  $N_{\min}(\ell+1) =N_{\min}(\ell)$  and
$N_{\max}(\ell+1)=N_{\max}(\ell)$,
we decided that the decay time was $100(\ell-1)$.
This procedure detects quiescent states with density 
periodicities up to 100, if any should occur.} 
In figure \ref{fig:decay1} we present the resulting $P_L(\tq)$.
Our results are fully consistent with the early work by
Bagnoli {\it et al.} \cite{Bagnolietal91} for lattices of up to $L=256$,
but present-day computational power allows for a much higher
precision. 
It appears that the distribution has a ``dead time'' during which
there is virtually zero probability for the system to reach its quiescent 
state, followed by a steep rise in this probability, which
quickly attains a maximum. 
Finally, as is clear from figure \ref{fig:decay1}, the curves decay
exponentially for large times.%
\footnote{For a periodic square lattice of $L=20$  it was determined by
Johnston \cite{Johnston09} that the
decay of the tail is exponential with very high accuracy.}
It appears that the decay times (that we may call $\tau_L^\prime$)
of the exponential tails are numerically indistinguishable from the $\tau_L$
obtained in figure \ref{fig:tauL}.
Hence we have from this simulation
\beq
P_L(\tq) \simeq a_L\,\ee^{-\tq/\tau_L}, \qquad \tq\to\infty,
\label{PLaspt}
\eeq
with decay times $\tau_L$ as in equation (\ref{xtauL}).

Bagnoli {\it et al.} \cite{Bagnolietal91}
considered the location $\tqm(L)$ 
where $P_L(\tq)$ peaks
and found that in the regime of system sizes they studied it may be
described by a power law $\tqm(L) \propto L^\zeta$ with $\zeta\approx 0.7$
(they denote this $\zeta$ by $z$).
In figure \ref{fig:decay4} we show their data points, 
as well as our own, 
for $L=128,\, 256$, and $512$.
For $L=128$ and $L=256$ our values are seen to
virtually coincide with those of Ref.\,\cite{Bagnolietal91}.
The data, however, suggest a downward curvature,
which is reinforced by our data point at $L=512$.
We investigate the large $L$ behavior of this curve in the next subsection.

\subsection{Heuristic argument}
\label{sec:theory}

We construct here for the curve $\tqm(L)$ of figure \ref{fig:decay4} 
a heuristic argument valid in the limit of asymptotically large $L$, 
which in practice is attained when $L\gtrsim \xi_\infty$.
In that limit we expect $\tqm(L)$ also to tend to infinity.

At a given time $t$, let us imagine 
an $L \times L$ lattice 
divided up into $L^2/\xi_t^2$ blocks of size $\xi_t \times \xi_t$.
Such blocks may be considered as statistically independent, 
due to not yet having had enough time to interact.
Let the function $c_L(t)$ indicate the probability at time $t$
that a $\xi_t\times\xi_t$ block be quiescent.
This function is unknown but we certainly expect it to increase with time
and be such that $c_L(\infty)=1$.
Let $Q_L(t)$ be the probability at time $t$ that an $L\times L$ system be
quiescent. For this to be true, it is necessary that all its
$\xi_t\times\xi_t$ blocks be quiescent, and therefore
\beq
Q_L(t) = c_L(t)^{L^2/\xi_t^2}.
\label{xQLt}
\eeq
Whereas mathematically $c_L(t)$ and $Q_L(t)$ are equivalent,
the tacit assumption here is that $c_L(t)$ is only weakly $L$ dependent,
and that we may exploit this feature.

In practice it is more convenient to work with
the function $f_L(t)$ defined by
\beq
Q_L(t) = \ee^{-L^2f_L(t)}, \qquad 
f_L(t) = \frac{1}{\xi_t^2} \log\frac{1}{c_L(t)}\,,
\label{dfLt}
\eeq
and which is also expected to be only weakly $L$ dependent.
Since $c_L(t) \to 1$ for $t\to\infty$, we have that $f_L(t)\to 0$
in that limit.

Using that $P_L(t) = (\dd/\dd t)\,Q_L(t)$ 
we get from equation (\ref{dfLt}) the expression
\beq
P_L(t) = -L^2f^\prime_L(t)\ee^{-L^2f_L(t)}.
\label{xPLt}
\eeq
The maximum of $P_L(t)$ is the solution of $(\dd/\dd t)P_L(t)|_{t=\tqm}=0$,
which gives 
\beq
f_L^{\prime\prime}(\tqm) = L^2f_L^{\prime 2}(\tqm).
\label{maxcond}
\eeq
We compare (\ref{xPLt}) to the findings of our simulation,
namely equation (\ref{PLaspt}),
and obtain after a time integration
\beq
f_L(t)= -\frac{1}{L^2}\,\log (1-a_L\tau_L\ee^{-t/\tau_L}),
 \qquad L\gg\xi_\infty\,.
\label{xfLt}
\eeq
In the large $L$ limit $\tau_L$ tends to $\tau_\infty\approx 1800$, 
and it appears from the simulation%
\footnote{This limit appears only when considering our last two curves,
the ones for $L=256$ and $L=512$.}
that the ratio $a_L/L^2$ tends to the fixed value
$a_L/L^2 \equiv A \approx 5.2\times 10^{-8}$. 
Upon inserting (\ref{xfLt}) in the maximum condition
(\ref{maxcond}) we obtain 
\beq
\tqm(L) = 2\tau_\infty \log L \, + \, \tau_\infty\log A\tau_\infty\,.
\label{soltqm}
\eeq
This curve, with the values of $A$ and $\tau_\infty$ 
as stated before, has been presented as the solid red line 
in figure \ref{fig:decay4}.
We see that 
for $L\gtrsim 256$ it is in excellent 
agreement with the two data points and provides a credible 
asymptotic expression for larger $L$. 

It should be noted, however, that this is a lowest order approximation,
based on the empirical input formula (\ref{PLaspt}).
We have not pursued the possibility of improving the result by
adding higher order correction terms to that formula.

\section{Discussion}
\label{sec:discussion}

We have considered the statistics of the Game of Life (GL)
on an $L\times L$ square lattice for sizes up to $L=512$.
Even though we used no special programming techniques,
our accuracy is higher than that of earlier work,
which mostly dates back one or two decades. 

We have studied finite size
effects and established the dependence of the asymptotic density
relaxation time $\tau_L$ on the system size $L$.
We found that in an intermediate range of system sizes
$\tau_L \propto L^z$ with a dynamical exponent $z=1.66 \pm 0.04$;
and that for system sizes $L\gtrsim 180$
the relaxation time $\tau_L$ saturates and approaches the constant value
$\tau_\infty=1800 \pm 50$, independent of system size.
 
We have performed the first study, to the best of our knowledge, of
correlation functions in the GL. We found that when the system
relaxes from a random initial state, 
the large distance decay of the pair density
may be characterized by a time dependent correlation
length $\xi_t$. We found that for an intermediate range of times
$\xi_t \propto t^{1/z^\prime}$ with $z^{\prime}$ close to $z$, 
whereas for times $t\gtrsim 6000$ there is saturation 
at a constant value $\xi_\infty=50 \pm 10$.

Hence there are lattice size independent cutoffs in space and time,
exactly as one would expect for a noncritical system.
Larger lattices have been considered by several investigators in the past,
but sizes larger than $L=512$ are not needed to reach our conclusions.
\vspace{2mm}

We briefly add a few more comments on the relation between our results
and other work 
that has been reported in the literature.

Bennett and Bourzutschky \cite{BennettBourzutschky91} obtain a correlation
length of $42 \pm 3$ lattice distances, which compares favorably with our
$\xi_\infty=50 \pm 10$. Their value is based, essentially, 
on the penetration length of the density into the system away from a 
boundary of cells kept alive randomly;
we have independently confirmed \cite{CornuHilhorstxx}
the length obtained by such a procedure.
The relaxation time of 
$200 \pm 10$ time steps reported by same authors \cite{BennettBourzutschky91}
is an {\it average\,} time and refers to
the relaxation of external perturbations; it cannot be
compared to the asymptotic time $\tau_\infty$ of present work.
\vspace{2mm}

We have not explored initial densities other than the value $\rhoin=0.3$.
It appears in simulations 
\cite{Bagnolietal91,Malarzetal98,Garciaetal93,Rozenfeldetal07}
that there is an interval $0.15 \lesssim \rhoin \lesssim 0.75$
for which the density $\rho^*$ of the final quiescent state is constant
within the accuracy of the simulation.
Gibbs and Stauffer \cite{GibbsStauffer97}, in particular,
starting from an initial density $\rhoin=0.5$,
obtained the same quiescent state density 
$\rho^*$ of our equation (\ref{xrhostar}) 
to within at least three decimals of accuracy. 
The curves $\rho_L(t)$ for different initial densities approach each other
rapidly (as it seems, exponentially fast in time).
This does not prove, but at least suggests,
that the exponents associated with this decay are universal with respect to 
$\rho_{\rm in}$ within the interval in question. 
We therefore speculate that the asymptotic relation between
the length and time scales found in the present work for $\rho_{\rm in}=0.3$
in fact holds in this whole interval of initial densities.

The question of universality may be asked also about the exponents $b$
(for the density; section \ref{sec:intermdecay}) and $\eta$ 
(for the correlation function; section \ref{sec:intermcorrdecay}).
Whereas $z$ and $z'$ concern the asymptotic exponential decay
(of the density in time, and of the correlation function
in space, respectively), the exponents
$b$ and $\eta$ refer to intermediate power law regimes.
Speculation, therefore, seems more dangerous here.
We have no data on $\eta$ for initial densities other than $\rho_{\rm in}=0.30$.
As far as $b$ is concerned, 
simulation of the density decay for different initial densities
shows that for $\rho_{\rm in}\lesssim 0.25$ and $\rho_{\rm in}\gtrsim 0.50 $
the power law regime becomes too ill-defined to extract an exponent
$b$; but that within these limit values there is no obvious
variation of $b$ with $\rho_{\rm in}$. 

We believe that the various questions touched upon in this discussion
leave much room for future research.


\section*{Acknowledgments}

The authors thank J.-M. Caillol for discussion and F. Bagnoli
for correspondence.


\appendix

\end{document}